\documentclass[english]{article}
\usepackage[T1]{fontenc}
\usepackage[latin9]{inputenc}
\usepackage{graphicx}
\usepackage{esint}

\makeatletter

\providecommand{\tabularnewline}{\\}

\makeatother

\usepackage{babel}
\begin{document}

\title{Nucleon polarizability and long range strong force from $\sigma_{I=2}$
meson exchange potential }

\author{Carl-Oscar Gullstr\"om, Andrea Rossi}

\date{18 july 2017}
\maketitle
\begin{abstract}
We present a theory for how nucleon polarizability may be used to
extract energy from nucleons by means of special electromagnetic conditions.
Also a new theory for a long-range strong force is introduced by enhancing
the role of the $\sigma_{I=2}$ meson in nucleon-nucleon potential
obtained through isospin mixed $\sigma$ mesons. The novelty in the
idea is to let an imaginary mass exchange particle be enhanced by
absorbing only one particle in an entangled state of two particles.
The imaginary mass particle is not intendent to be free and contravene the laws of physics; it is merely included as a binding exchange particle
in a system with total positive invariant mass. In order to validate
part of the theory, we introduce an experiment that in many ways has
motivated this study.
\end{abstract}

\section*{Introduction}

A theory on the possibility that nuclear effects can occur at energies
lower than expected comes with two problems:
\begin{itemize}
\item How to extract energy without strong radiation.
\item How to effect this over a long range, i.e. ranges found by atomic separation
of nuclides instead of separation of nucleons in a nuclide. 
\end{itemize}
Two important assumptions are made in this paper to answer these questions.
The first is that a special electromagnetic field can create a substitute
for the meson exchange potentials created by nucleon-nucleon interaction
inside nuclides. This is motivated, f.i., by the meson-photon couplings
found in hadronic light by light scattering. The second assumption
is that the same special EM fields are capable of creating an environment
where nucleons may be transferred between nuclides

To match the theoretical predictions, links to experimental results
should be established. Such possible experimental results are listed
in the appendix. 

There are two relevant experimental results worth mentioning. The
first is that energy is released in the form of heat and kinetic energy
of heavy ions. The second result is that during the process nucleons
are transferred between nuclides causing chemical and isotopic shifts.

Our proposed theory consists of four parts. The first is an interpretation
of nucleon polarizability on how energy is extracted from theoretical
formulas and parameters of polarizability. The purpose of this theory
is to find which special electromagnetic field conditions correspond
to the extraction of energy out of the nucleon. The second part is
a short summary of the links between polarizability and nucleon-nucleon
(N-N) interaction indicating how polarizability interaction sets the
nucleon in a state that is found in nuclides.

The third part contains a new idea developed to answer the second
question, i.e. how to conduct nucleon transfer over a long-range.
The idea is that in meson exchange potential-based nucleon-nucleon
interaction there is a degenerated state between two scalar mesons
differing only in their isospin state. From $\pi\pi$ scattering,
two resonances at $36.77m_{\pi}^{2}$ and $-21.62m_{\pi}^{2}$ are
found(henceforth referred to as $\sigma_{I=0}$ and $\sigma_{I=2}$).
The second resonance does not correspond to a real particle (because
of its negative energy) and is only a parameter used in formulas for
interaction. In N-N interaction these states are mixed at a resonance
mass of 550 MeV. The idea is that, by enhancing interaction with isospin
0 particles (electrons and photons), the exchange meson changes its
mass parameter to a value close to zero, giving it the same interaction
length as the photon. The process is similar to that of enhancing
the $K_{L}$ over $K_{s}$ meson parts of $K_{0}$ by performing measurements
a long time after the creation of the meson, or choosing a proton
over a neutron by defining a specific charge in a strong force calculation
which would otherwise be charge-independent.

The fourth part describes possible atomic sources for the special
electromagnetic field conditions found from polarizability. Unfortunately
the available experimental data is not precise enough to draw any
definite conclusions here, hence only suggestions are made. 

To solve the two main problems listed above, the first thing to do
is to set the low energy scale. Low energy means that there is not
enough energy to push nucleons close enough together to interact strongly
with one another. The residual interaction is the electromagnetic
interaction with photons and electrons. For a single nucleon, the
theory resorted to here is nucleon polarizability, i.e. how the internal
structure of the nucleon changes with the interaction of photons.
From the standpoint of polarizability the first question may be answered
in two ways. Spin polarizability does not correspond to a classic
electromagnetic field, and therefore all momentum transfers from the
interaction should be released as kinetic energy of the nucleon. The
other solution is that there is a magnetic quadrupole state in the
nucleon with an energy level lower than zero. The full transition
to this state is not possible through energy conservation, yet it
is possible to extract parts of the energy over a limited time. The
solution to the long-range question is also twofold. The first solution
is the one described in the third part of the theory. The second is
a less plausible idea that is assisted by the special electromagnetic
field. The e-N interaction strength is in the same range as N-N, since
the electron replaces the full attractive potential corresponding
to a nuclide.

\section*{Nucleon Polarizability}

The main theory for nucleon polarizability resorted to here is the
baryon chiral perturbation theory \cite{Lensky}. Other extant theories
are the heavy baryon chiral perturbation theory and the fixed-t dispersion
theory, both of which have polarizability constants within the same
ranges as those of the baryon ChPT. For a review, see for instance
\cite{Pol rev}. The theory of polarizability is carried out by first
taking the ground state $E_{0}$ of the nucleon and then perturbing
the N-$\gamma$ interaction with an effective Hamiltonian($H_{eff}$
):

\[
H=E_{0}-H_{eff}
\]

If the condition $H_{eff}<0$ is fulfilled, the new state is an excitation
and $H_{eff}>0$ corresponds to a binding energy which could be used
to extract energy out of the nucleon. The effective Hamiltonian admits
an expansion of the form $H_{eff}=\sum H_{eff}^{(i)}$, where i denotes
the number of space time derivatives of the electromagnetic field
$A_{\mu}\left(x\right)$. According to reference \cite{Heff org 2},
to the fourth derivative $H_{eff}^{(i)}$ is given by:

\begin{equation}
\begin{array}{c}
H_{eff}^{\left(2\right)}=-\frac{1}{2}4\pi\left(\alpha_{E1}\bar{E}^{2}+\beta_{M1}\bar{H}^{2}\right)\\
H_{eff}^{\left(3\right)}=-\frac{1}{2}4\pi\left(\gamma_{E1E1}\bar{\sigma}\cdot\left(\bar{E}\times\dot{\bar{E}}\right)+\gamma_{M1M1}\bar{\sigma}\cdot\left(\bar{H}\times\dot{\bar{H}}\right)-2\gamma_{M1E2}E_{ij}\sigma_{i}H_{j}+2\gamma_{E1M2}H_{ij}\sigma_{i}E_{j}\right)\\
H_{eff}^{\left(4\right)}=-\frac{1}{2}4\pi\left(\alpha_{E1\nu}\dot{\bar{E}}\,^{2}+\beta_{M1\nu}\dot{\bar{H}}\,^{2}\right)-\frac{1}{12}4\pi\left(\alpha_{E2}E_{ij}^{2}+\beta_{M2}H_{ij}^{2}\right)
\end{array}\label{eq:heff}
\end{equation}

where $\alpha_{x}$, $\beta_{x}$, $\gamma_{x}$ are polarizability
constants, E and H are components of the electromagnetic fields. $\sigma$
is the Pauli spin matrices of the nucleon and $E_{ij}$ is given by
$E_{ij}=\frac{1}{2}\left(\nabla_{i}E_{j}+\nabla_{j}E_{i}\right)$
with the same relation for $H_{ij}$ where i and j stand for space
indexes. For the polarizability constants with an even number of perturbations,
the experimental values and theoretical predictions are shown in table
\ref{tab:alphabeta}. For the effective hamiltonians with an even
number of space time derivates every part of the EM field is a square;
there the only time the condition $H_{eff}>0$ is fulfilled is when the
polarizability constant is negative. In this case only theoretical
values exist: the polarizability constant of the magnetic quadrupole
and the electric dispersive polarizability constant. 

The odd number of perturbation is called spin polarizability and has
no meaning in a classic EM field. This means that the interaction
is due to the non standard electromagnetic parts of the nucleon, i.e.
the strong force. The situation is a bit more complicated if one wants
to find the conditions for $H_{eff}>0$. Since the field includes
the nucleon spin, the positive value condition depends on the alignment
of the nucleon spin with the EM field parts. An interesting and experimentally
easily accessible part of the polarizability is the case of forward
and backward scattering. These polarizability constants are labeled
$\gamma_{0}$ and $\gamma_{\pi}$. They are related to the spin polarizability
constants by: 

\[
\gamma_{0}=-\gamma_{E1E1}-\gamma_{E1M2}-\gamma_{M1M1}-\gamma_{M1E2}
\]

\[
\gamma_{\pi}=-\gamma_{E1E1}-\gamma_{E1M2}+\gamma_{M1M1}+\gamma_{M1E2}
\]

The spin polarizability experiment and theoretical values are displayed
in table \ref{tab:gamma}. The binding conditions are fulfilled for
the proton, for example by the $\gamma_{0}$ constants; and there
is a possibility to define regions where $H_{eff}>0$ when the four
spin polarizability constants are known.

\begin{table}
\begin{tabular}{|c|c|c|c|c|c|c|}
\hline 
 & $\alpha_{E1}$ & $\beta_{M1}$ & $\alpha_{E2}$ & $\beta_{M2}$ & $\alpha_{E1\nu}$ & $\beta_{M1\nu}$\tabularnewline
\hline 
\hline 
Proton &  &  &  &  &  & \tabularnewline
\hline 
B$\chi$PT Theory\cite{Lensky} & $11.2\pm0.7$ & $3.9\pm0.7$ & $17.3\pm3.9$ & $-15.5\pm3.5$ & $-1.3\pm1.0$ & $7.1\pm2.5$\tabularnewline
\hline 
Experiment(PDG\cite{PDG}) & $11.2\pm0.4$ & $2.5\pm0.4$ &  &  &  & \tabularnewline
\hline 
Neutron &  &  &  &  &  & \tabularnewline
\hline 
B$\chi$PT Theory\cite{Lensky} & $13.7\pm3.1$ & $4.6\pm2.7$ & $16.2\pm3.7$ & $-15.8\pm3.6$ & $0.1\pm1.0$ & $7.2\pm2.5$\tabularnewline
\hline 
Experiment(PDG\cite{PDG}) & $11.8\pm1.1$ & $3.7\pm1.2$ &  &  &  & \tabularnewline
\hline 
\end{tabular}

\caption{\label{tab:alphabeta}Theoretical and experimental values of the proton
and neutron static dipole, quadrupole and dispersive polarizabilities.
The units are $10^{-4}$ $fm^{3}$ (dipole) and $10^{-4}$ $fm^{5}$
(quadrupole and dispersive). }
\end{table}

\begin{table}
\begin{tabular}{|c|c|c|c|c|c|c|}
\hline 
 & $\gamma_{E1E1}$ & $\gamma_{M1M1}$ & $\gamma_{E1M2}$ & $\gamma_{M1E2}$ & $\gamma_{0}$ & $\gamma_{\pi}$\tabularnewline
\hline 
\hline 
Proton &  &  &  &  &  & \tabularnewline
\hline 
B$\chi$PTTheory\cite{Lensky} & $-3.3\pm0.8$ & $2.9\pm1.5$ & $0.2\pm0.2$ & $1.1\pm0.3$ & $-0.9\pm1.4$ & $7.2\pm1.7$\tabularnewline
\hline 
MAMI 2015\cite{Mami2015} & $-3.5\pm1.2$ & $3.16\pm0.85$ & $-0.7\pm1.2$ & $1.99\pm0.29$ & $-1.01\pm0.13$ & $8.0\pm1.8$\tabularnewline
\hline 
Neutron &  &  &  &  &  & \tabularnewline
\hline 
B$\chi$PT Theory\cite{Lensky} & $-4.7\pm1.1$ & $2.9\pm1.5$ & $0.2\pm0.2$ & $1.6\pm0.4$ & $0.03\pm1.4$ & $9.0\pm2.0$\tabularnewline
\hline 
\end{tabular}

\caption{\label{tab:gamma}Theoretical and experimental values of the proton
and neutron static spin polarizabilities. The units are $10^{-4}$
$fm^{4}$. }

\end{table}

\subsubsection*{Discussion of the condition $H_{eff}>0$ }

The polarizability condition $H_{eff}>0$ is strange from the wiev
of energy conservation. The condition needs some special attention
to explain why it does not violate any basic law of physics. To understand
the states with positive values of $H_{eff}$ a thermodynamic view
could be implemented. Start with the first law of thermodynamic:

\[
dU=dQ+dW
\]

where $dU$ is the change in interanal energy, $dQ$ the thermal flow(assumed
to be 0 here) and $dW$ the internal work of the system. For the groundstate
particles, the proton and photon the internal energy U would be defined
as 0 if there is no internal kinetic energy. In the case when $E_{\gamma}\sim0$
which is the level where the static polarizability constant is defined
the condition $dU>0$ has to be fullfilled. With a large $E_{\gamma}$
the polarizability interaction would be called dynamic and the equation
\ref{eq:heff} would no longer be valid. For small magnetization and
electric polarizability, the dW is defined by:

\[
dW=-\mu_{0}VHd\left(\chi H\right)
\]

\[
dW=-\varepsilon_{0}VEd\left(\left(\varepsilon_{r}-1\right)E\right)
\]

where V is the volume of the system. The condition $\chi>0$ and $\varepsilon_{r}>1$
would be forbidden by the $dU>0$ condition. This would correspond
to the situation $H_{eff}>0$ in polarizability calculations. The
solution to the $H_{eff}>0$ problem is the definition of $U=0$.
If the proton is considered the ground state, then the internal energy
is 0; but if the bound nucleon in a nuclide is considered the G.S.,
then there is an internal energy that is defined by the mass difference.
This gives a second problem, i.e. why is there no spontaneous de-excitation
of the nucleon. A basic example is why the process $p+n\to d+\gamma$
is allowed but not $p\to p*\left(d\right)+\gamma$. The solution is
different for the spin polarizability and quadrupole polarizability.
For spin polarizability, a solution is that the interaction is due
to non-classic EM fields. The basic interpretation of this is that
the interaction cannot form real photons. Virtual photons are still
possible, but the reaction $p\to p*+\gamma*$ is a normal interaction
found for particles, which is a source for a static EM field. For
the quadrupole polarizability the solution to the forced forbidden
spontaneous de-excitation is different. The only solution would be
that the hypothetical ground state have negative energy, i.e. the
full reaction $p\to p*+\gamma$ is forbidden by energy conservation.
This is an expected property of the nucleons as long as they not
decay. The upper limit of proton decay is of the order of $10^{31}$
years, according to one of the best-measured upper limits on Earth.
The reason this is expected comes from the properties of the internal
structure of the proton. The internal structure through EM interaction
consists of three charged quarks with no electric dipole moment. The
electric structure then consist of two perfect aligned opposite dipoles.
The opposite dipoles are capable of forming two photons, so that spontaneous
de-excitation $p\to e^{+}+2\gamma$ would be possible if one assumes
that lepton and baryon number could be exchanged.

A state with energy lower than zero is also found when expanding the
instance of spontaneous de-excitation from photons only to photons
plus mesons. The $\sigma_{I=2}$ meson has a mass square below 0. The
spontaneous de-excitation would then be $p\to p*+\sigma$ instead
of $p\to p*+\gamma$ but the $\sigma$ would stay in place by the
absence of energy available for kinetic energy separation of the meson
and proton.

\subsection*{Identifying the Special EM Field}

In this section we give criteria for the spin polarizability and the
magnetic quadrupole polarizability to meet the condition$H_{eff}>0$.
In order to do so the EM field condition may be divided in two parts.
The first is when the system is an isolated $\gamma-N$ system. This
is the case in which the polarizability constants are theoretically
calculated: here, limits on the EM fields are dictated by the symmetry
condition of the total wave function. The second part is when there
exists at the same time an extra particle/ weak-coupled EM field,
which would allow the full range and direction of each parameter.
With the allowed full parameter range, there is one more important
condition that has to be fulfilled. The $H_{eff}>0$ condition has
to be set for the full interaction, not just parts of the field. 

Starting with the interaction with $\bar{E}$ and $\dot{\bar{E}}$
the full $H_{eff}$ could be divided in two parts, either with $\dot{\bar{E}}$
aligned longitudinally or transversally to $\bar{E}$. Introducing
the new paramter $x_{L,T}=\dot{\bar{E}}_{L,T}/\bar{E}\cdot\hat{\bar{E}}/\hat{\dot{\bar{E}}}$,
two second order equation arise:

\begin{equation}
\alpha_{E1}\pm\gamma_{E1E1}x_{T}+\alpha_{E1\nu}x_{T}^{2}\label{eq:}
\end{equation}

\begin{equation}
\alpha_{E1}+\alpha_{E1\nu}x_{L}^{2}\label{eq:-1}
\end{equation}

where the $\pm$ sign depends on the direction of the nucleon spin
vector compared to $\left(\bar{E}\times\dot{\bar{E}}\right)$. By
solving the two second order equations, ranges of x where $H_{eff}>0$
is fulfilled could be defined. The ranges are displayed in table \ref{tab:Ranges-of-doteE}
for the theoretical values of the polarizability constants.

\begin{table}
\begin{tabular}{|c|c|c|}
\hline 
Nucleon & $\bar{\sigma}\cdot\left(\bar{E}\times\dot{\bar{E}}\right)$ & $x=\dot{\bar{E}}/\bar{E}$ range (fm)\tabularnewline
\hline 
\hline 
p & + & $x_{T}<-2\; x_{T}>4.5$\tabularnewline
\hline 
p & - & $x_{T}<-4.5\; x_{T}>2$\tabularnewline
\hline 
p & 0 & $x_{L}^{2}>0.11$\tabularnewline
\hline 
n & + & $3.1<x_{T}<44$\tabularnewline
\hline 
n & - & $-44<x_{T}<-3.1$\tabularnewline
\hline 
n & 0 & -\tabularnewline
\hline 
\end{tabular}

\caption{\label{tab:Ranges-of-doteE}Ranges of $\dot{\bar{E}}/\bar{E}$ where
the condition $H_{eff}$ is fullfilled for the nucleons.}

\end{table}

For the magnetic field components $\bar{B}$ and $\dot{\bar{B}}$
the $H_{eff}>0$ condition is never fullfilled. The condition on the
polarizability constants to allow for this situation would be(with
positive $\beta_{M1}$ ):

\begin{equation}
\frac{\gamma_{M1M1}^{2}}{4\beta_{M1\nu}}>\beta_{M1}\label{eq:-2}
\end{equation}

which is not supported by the theoretical values from B$\chi$PT. 

In the case of spin polarizability combined with electric quadrupoles,
there is also no region where $H_{eff}$ is larger than 0. The x parameter
introduced here for a second order equation would be $x=\sigma_{i}H_{j}/E_{ij}$(with
E and H changed for magnetic quadrupole), with the maximum for the
spin polarizability part occuring when the magnetic field is completely
in space dimension j. The condition for negative values is now:

\begin{equation}
\frac{6\gamma_{E2M1}^{2}}{\alpha_{E2}}>\beta_{M1}\label{eq:-3}
\end{equation}

which is not fullfilled for the theoretical values. For the magnetic
quadrupole the negative sign of the magnetic polarizability constant
changes the condition to allow for a large region. Defining $E_{j}$
as $Esin\theta$(with $\theta$ the angle between dimension j and
the plane defined by i and k) gives the second order equation:

\begin{equation}
\beta_{M2}/6+2sin\theta\gamma_{E1M2}x+x^{2}\alpha_{1}\label{eq:-4}
\end{equation}

The condition for having $H_{eff}$=0 values are fullfilled by:

\begin{equation}
\frac{6\gamma_{E1M2}^{2}}{\beta_{M2}}sin\theta=\alpha_{E1}\label{eq:-5}
\end{equation}
For $\theta$ above this value, an unwanted $H_{eff}<0$ region exist
within the solutions to the second order equation.

The three conditions for $H_{eff}>0$ is therefore a center of a magnetic
quadrupole(eq. \ref{eq:-4}) which also allows for a weak electric
field and two ranges from the parameter $\dot{\bar{E}}_{L,T}/\bar{E}$(eq.
\ref{eq:} and \ref{eq:-1}).

\newpage{}

\section*{Polarizability and nucleon-nucleon interactions}

\begin{figure}
\includegraphics[scale=0.5]{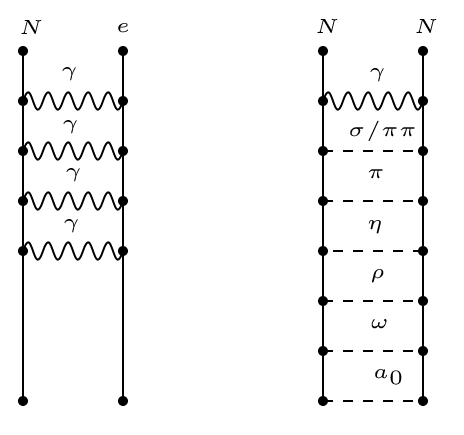}\caption{\label{fig:Left:-Polarizability-from}Left: The polarizability is
the interaction of a nucleon with one or many, real or virtual photons.
Here with an electron as the source of the photons. Right: Besides
the photon nucleon-nucleon interaction also includes mesons.}

\end{figure}

\begin{figure}
\includegraphics[scale=0.33]{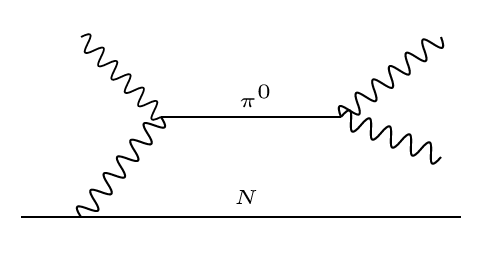}\includegraphics[scale=0.33]{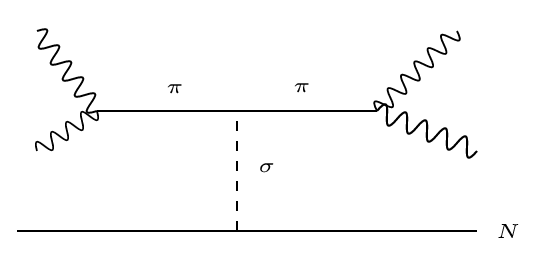}\caption{\label{fig:Left:-Primakoff-effect.}Left: Primakoff effect. Right:
Instead of having a photon from the nucleon, a $\sigma$ meson is between
the nucleon and the photons. Here with pions in-between, even if direct
photon $\sigma$ interaction exist. }
\end{figure}

Since polarizability is a temporary interaction state extracting energy
out of the nucleon, the state is therefore also temporary. Permanent
energy extraction out of nucleons could be found in fusion of nucleons
into nuclides (and transfer of nucleon between nuclides). Therefore,
to make the temporary energy extraction permanent, one needs to move
the polarizability interaction into nucleon-nucleon interaction. To
do this, a link between nucleon- nucleon interaction and $\gamma$/e-nucleon
interaction has to be established, to see that the polarizability
interaction sets the nucleon in the same state it has when it is bound
in a nuclide. 

Figure \ref{fig:Left:-Polarizability-from} visualizes the difference
between polarizability and nucleon-nucleon interaction in terms of
exchange particles. To establish a link between the two, one has to
introduce meson-photon couplings. This is, for example, used in hadronic
light by light scattering and Primakoff effect. However, since the
goal is the binding interaction, the interaction particle with a source
in the nucleon should not be a photon but a $\sigma$. Figure \ref{fig:Left:-Primakoff-effect.}
shows a comparision of Primakoff and $\gamma-\sigma-N$ interaction.
The link is established by finding an alignment in the two types of
interaction due to states with the same quantum numbers, i.e. parity,
spin and charge should be conserved. Nucleon-nucleon interaction is
suppressed or enhanced compared to polarizability, depending on the
extra coupling constants between the mesons and the photon.

A second motivation for the link is to establish a space range of
polarizability. If polarizability uses the exchange particles of the
strong force(which is motivated by the absence of spin polarizability
in a classical EM field), one would expect the interaction range to
be similar to that of the strong force.

From the first attempt to accomplish the Yukawa nucleon-nucleon interaction
\cite{Nuclear binding theory base(yukawa)} using a pion as a heavy
exchange particle, the best known today nucleon-nucleon interactions
are obtained by using quantum Monte Carlo methods \cite{AV18illsum}.
The operators used there are not only meson exchange but also $\Delta$
baryons. 2N and 3N forces with more than one exchange particle are
included and fitting is done by the Monte Carlo method. Each term
in the Hamiltonian is a multiplication of operators including space,
spin, isospin, and orbits of the nucleons with a fitted coupling strength.
The polarizability binding link for these operators is found; it is
equal to the one corresponding to the exchange that gives $H_{eff}>0$
for polarizability and a binding term in the Hamiltonian of the N-N
interaction. Other interaction models exist, for example the Bonn
model\cite{bonnnuclbinding} is further described with one boson exchange,
and links each operator with a meson exchange. In this model the fitted
parameters are coupling constants for each meson instead of each operator
set.

The condition $H_{eff}>0$ from polarizability is met for the magnetic
quadrupole polarizability when magnets are opposite and separated
on the z axis. In operator form this means that $\bar{s}\cdot\bar{r}$
is nonzero. A match for the binding polarizability constants are the
tensor and spin orbit operators. In the tensor operator the scalar
product of space and total spin is non zero. The operator used is
$S_{12}=2\left(\frac{3\left(\bar{s}\cdot\bar{r}\right)}{r^{2}}-\bar{s}\cdot\bar{s}\right)$
which in a one boson exchange model corresponds to $\eta$ exchange.
$S_{12}$ is non zero only when the spin-spin couplings are in triplet
state; a magnetic quadrupole is then only present when $s_{z}$ is
opposite. The second operator is the spin orbit coupling L{*}S which
corresponds to $\sigma$ meson exchange. This operator could have
a back-to-back magnet with one of the magnets coming from the spin
and the other from the orbit. It is notable that the $\sigma$ meson
is not a well-established particle, and that this interaction is usually
described as an s-wave scattering of two pions. For the binding terms
of dispersive polarizability the operator needs to include a time
operator. That is also found in spin orbit coupling since the L terms
include a time derivative.

\section*{Long-range strong force}

\begin{figure}
\includegraphics[scale=0.4]{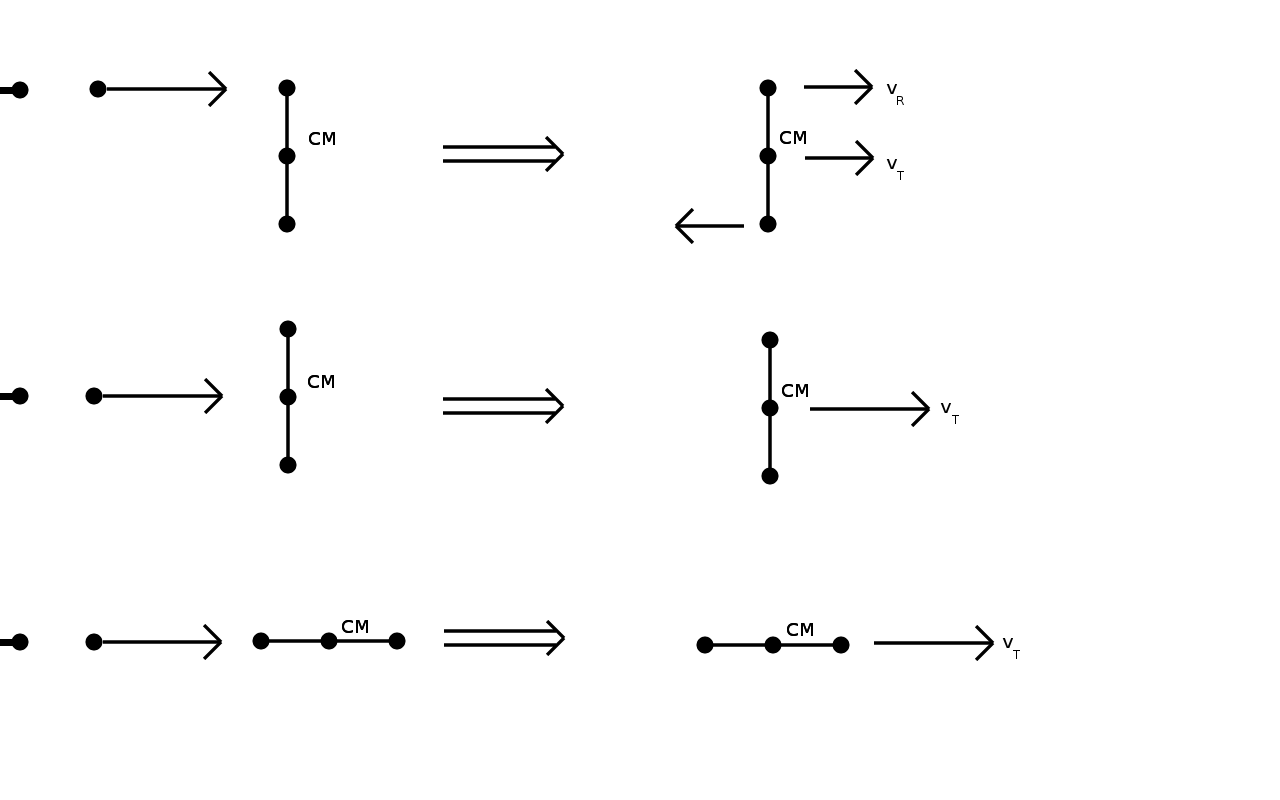}\caption{\label{fig:figure of the gamma problem}When two particles fuse to
one, and the target has internal structure, a hit at the center of
mass would create translation motion only, while a not direct hit would
create both translation and circular motion. Note that in order to
conserve energy and momentum a source particle must be present to avoid
the creation of a new particle. }

\end{figure}

\begin{figure}
\includegraphics[scale=0.5]{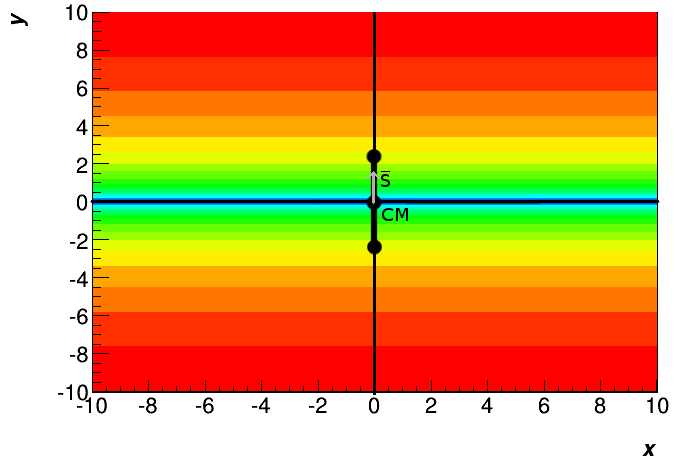}

\caption{\label{fig:potential expamles}x-y view of a potential formed by $\bar{s}\cdot\bar{r}$
with $\bar{s}$ on the y axis together with line potentials x=0 and
y=0(black lines).}

\end{figure}

Putting a nucleon in the special EM condition $H_{eff}>0$ could extract
energy out of it but the effect is only temporary. When the nucleon
gets out of place in a normal neutral EM environment, polarizability
would absorb energy by scalar polarizability up to the ground state.
To transfer nucleons over a long-range, a long-range potential of
the strong force has to be established.

A less probable alternative to the long-range potential is if the
e-N coupling in the special EM field environment would create a strong
enough binding to compare an electron with a full nuclide. In this
hypothesis, no constraints on the target nuclide are set, and nucleon
transition to excited states in the target nuclide should be possible.

In other words these two views deals with the electrons role. One
is as a carrier of the nucleon and the other is as a trigger for a
long-range potential of the nucleon. The motivation for the second
case is visualized in figure \ref{fig:figure of the gamma problem}.
A de-exitation $N*\to N+\gamma$ would require an internal EM oscillation
motion in N{*}. This oscillation would only occure if the transfered
nucleon would hit the target nuclide outside the center of mass. Examples
of the special potential are found in figure \ref{fig:potential expamles}
where the example potential with a $\bar{s}\cdot\bar{r}$ term enhances
the chance for a central hit, while the example potential x=0 and
y=0 only allows for a direct hit. A line potential is formed from
when a dot and crossproduct is combined like in the case $\sigma\cdot\left(\bar{E}\times\dot{\bar{E}}\right)$
and $\sigma\cdot\left(\bar{B}\times\dot{\bar{B}}\right)$ that could be found in
polarizability.

The nucleon-nucleon interaction is short-ranged because of massive
exchange particles. The basic short range from Yukawa one-pion exchange
potential\cite{Nuclear binding theory base(yukawa)} is given(without
spin and isospin dependency) by:

\[
\frac{g}{4\pi}\frac{e^{-m_{\pi}r}}{r}
\]

where $m_{\pi}$ is the pion mass and g is a Yukawa coupling. This
is to be compared to the electromagnetic potential:

\[
\frac{q}{4\pi}\frac{e^{-m_{\gamma}r}}{r}
\]

where q is the charge of the source particle and $m_{\gamma}=0$ gives
the Coulomb potential with space dependency of $1/r$ .

A long-range strong force is found if an exchange particle has a mass,
or rather a squared mass, equal or less than the photon mass:

\[
m^{2}\leq m_{\gamma}^{2}
\]

Notable here is that the mass should only be used in a system where
the total mass is positive, i.e. as a particle exchange term in a total
two particle system. Such a meson is found in $\pi\pi$ s-wave scattering
in the isospin 2 channel. In nucleon-nucleon interaction $\pi\pi$
s-waves are found in 3N forces. The pion s-wave from the Illinois
model is not accurate and only a plausible 1MeV size strength is included,
but it is needed to explain energy levels of light nuclides. To move
to the one boson exchange Bonn model the $\sigma$ exchange potential
is there given by:

\[
V_{NN}^{\left(\sigma\right)}\left(r\right)=\int\frac{d^{3}q}{\left(2\pi\right)^{3}}e^{iqr}\frac{g_{\sigma NN}^{2}}{-q^{2}-m_{\sigma}^{2}}=-\frac{g_{\sigma NN}^{2}}{4\pi}\frac{e^{-m_{\sigma}r}}{r}
\]

The theory of $\pi\pi$ scattering may be found in \cite{eff theory both sol}.
This paper uses both fixed-t dispersion and $\chi$PT as seen in $B\chi$PT
polarizability theory. In $\pi\pi$ scattering the $\sigma$ mass
is related to scattering phase shifts. The phase shift equation is
\cite{eff theory phase shifts}: 

\[
tan\delta_{l}^{I}=\sqrt{1-\frac{4m_{\pi}^{2}}{s}}q^{2l}\left\{ A_{l}^{I}+B_{l}^{I}q^{2}+C_{l}^{I}q^{4}+D_{l}^{\hat{I}}q^{6}\right\} \left(\frac{4m_{\pi}^{2}-s_{l}^{I}}{s-s_{l}^{I}}\right)
\]

where q is the momentum transfer of the scattering pion, A...D are
constants, l gives spin and I isospin channel. The kinematic variable
s is given by $s=4m_{\pi}^{2}+q^{2}$ and $s_{l}^{I}$ specifies where
$\delta$ passes through $90^{o}$. At lowest order the A constants
equal the s-wave scattering length $a_{0}^{0}=0.22$ and $a_{0}^{2}=-0.044$
\cite{eff theory phase shifts}. The negative sign of $a_{0}^{2}$
is what gives the necessary $m^{2}\leq m_{\gamma}^{2}$ relation.
The parameter $s_{l}^{I}$ is given by $s_{0}^{0}=36.77m_{\pi}^{2}$,
$s_{1}^{1}=30.72m_{\pi}^{2}$ and $s_{0}^{2}=-21.62m_{\pi}^{2}$.

The isospin 1 channel is isolated in a L=1 state by Dirac motivation
of a total antisymmetric wave function, and this pole corresponds
to the $\rho$ meson. Since isospin is not a perfect symmetry the
L=0 channel is mixed and $\sigma$ exchange in nucleon nucleon interaction
includes both I=0 and I=2 $\sigma$ mesons. The $m_{\sigma}$ used
in the Bonn model is 550 MeV while $s_{0}^{0}$ gives a mass at 844
MeV. Assuming an equal mix of $\sigma_{I=2}$ and $\sigma_{I=0}$ in
$NN\sigma$ coupling gives the more correct mass of $m_{\sigma}=$543
MeV.

To create a pure potential with $\sigma_{I=2}$ meson only the $\sigma_{I=0}$
has to be absorbed. Since the electron does not have an isospin component,
e-N binding through nucleon polarizability is reducing the $\sigma_{I=0}$
part of the mixed $\sigma$ exchange meson. To get a size of the free
$\sigma_{I=2}$ interaction one could start with $\gamma$-vector
meson couplings. In the quark model the coupling ratio is given by
$g_{\omega\gamma}=3g_{\rho\gamma}$ and $g_{\phi\gamma}=-\frac{3}{\sqrt{2}}g_{\rho\gamma}$.
Assuming that two vector mesons are the strongest part of pion coupling,
one gets a factor of two, while $\sigma$ from $\pi\pi$ scattering is
also a factor of 2. The total enhancement of $\sigma_{I=2}$ potential
in e-N coupling is a factor of $\left(g_{\omega\gamma}/g_{\rho\gamma}\right)^{4}$,
i.e. $3^{4}=81$. Using this suppression and assuming equal mix of
isospin states, the $\sigma$ mass becomes virtual with the value
$-im_{\sigma}=$ 642 MeV. The $\sigma$ exchange s-wave scattering
then changes the exponential term to an oscillating one. Since the
mass(square) of $\sigma_{I=2}$ is negative and the total mass of
the nucleon should be positive and given by the sum of scalar particles,
the potential is directed straight into the center of mass of the
nucleon.

Since isospin comes with direction $I_{z}$ that equals charge($q=1/2-I_{Z}$),
the type of $\sigma_{I=2}$ matters for the potential. Proton has
$I_{Z}=+1/2$ and neutron $I_{Z}=-1/2$ while $\Delta$ has $I_{z}=\pm3/2$
and $I_{z}=\pm1/2$ depending on the charge . From opposite attract
properties of the potentials, protons are attracted to potentials
created by neutrons and neutrons to potentials created by protons.
To create a full $\sigma_{I=2}$ meson N$\Delta$e is enhanced, compared
to NNe where the $\Delta$ comes from the 3N force i.e. 3N-e interaction
is most probable needed. In full 4N ground state clusters the N-N
potential fulfills the binding polarizability conditions and e/$\gamma$-N
interaction can not extract energy. This leaves 1 hole nuclides as
the only source of $\sigma_{I=2}$ long range potentials. Examples
of isotopes are $^{7}Li$, $^{27}Al$ and in the case of attractive
3 neutron cluster $^{61}Ni$. Since most free 3N states have 2 neutrons
and 1 proton (except $^{3}He$) the majority of $\sigma_{I=2}$ potentials
are attractive to protons. Neutron attractive $\sigma_{I=2}$ potentials
could be formed if a proton transfer does not form a perfect 4N state.
For example the reaction Ni+p{*} with p{*} from manganese or lithium
would give a copper isotope below the ground state. Such a reaction
would still be possible as a temporary unstable state with the aid
of a proton attractive $\sigma_{I=2}$ potential from for example
a 3N cluster in $^{27}Al$. Neutron transfer should be stronger compared
to proton transfer since there is no Coulomb repulsion between the
proton and the target nuclide.

\section*{}

\subsection*{Energy distribution between electron and nuclides}

The ideas for energy release in this paper are so far two. Polarizability
interaction requires kinetic energy to be shared between an electron
(or more) and a nuclide i.e. $E_{rel}=E_{k}\left(e^{-}\right)+E_{k}\left(N_{source}\right)$,
while nucleon transfer between nuclide $N_{1}$ and $N_{2}$ to nuclide
$N_{3}$ and $N_{4}$, suggests that the mass difference between the
start and final nuclides are shared between the two nuclides. From
the center of mass source of the long range potential only kinetic
energy is created i.e. $E_{rel}=E_{k}\left(N_{3}\right)+E_{k}\left(N_{4}\right)$.
A mix between the two are expected i.e.:

\begin{equation}
E_{rel}=E\left(N_{1}+N_{2}\right)-E\left(N_{3}+N_{4}\right)=E_{k}\left(e^{-}\right)+E_{k}\left(N_{3}\right)+E_{k}\left(N_{4}\right)\label{eq:erel}
\end{equation}

There is no experimental data to set any numbers into this equation.
In the case that the electron works as a pure trigger the condition
$E_{k}\left(e^{-}\right)=0$ is fullfilled. The condition $E_{k}\left(N_{1}\right)+E_{k}\left(N_{2}\right)=0$
is less probable since this would require a spot on energy release
from the electron nuclide polarizability interaction.

\section*{Creating the special EM field}

Before the $\sigma_{I=2}$ potential has been established the special
EM fields has to come from short range interaction by the link to
nucleon-nucleon interaction. The source of the fields is then enhanced
by atomic sources over external fields. The best source are S-state
electrons since they have an non-zero component of the wave function
at r=0.

The two polarizability conditions have various sources. The ranges
from $x_{T}=\dot{\bar{E}}_{T}/\bar{E}$ are fullfilled for an electron
orbiting the nucleon in an circular motion. This is because in a perfect
circular motion $\dot{\bar{E}}_{T}$ is constant and always perpendicular
compared to $\bar{E}$. However the $x_{T}$ ranges are above 1 fm(approximately
the radius of the circular motion), and atomic sources would have
a radius of 1\AA. Here we propose three theoretical ideas to overcome
this problem. The first one is to use high pressure to shrink the
atoms to a small enough size. The second is to use a current. For
a nuclide with a nucleon that has a stable distance to the charge
center, the S-state electrons would temporarily be viewed as a part
of a rotation when passed through the charge center. However, the
part circular motion is opposite when the electron passes through
the next time in the oscillating motion. But if the electron motion
were a current following the S-state path, the return part of the
oscillation would never be fulfilled. The third option is when the
relative direction between the nucleon and the charge center of the
nuclide oscillates with the same speed as the S-state oscillation.

For magnetic quadrupole polarizability, the electron alone can not
form a quadrupole, which means that an external magnetic field would
be needed. To have a quadrupole would require that the magnetic moment
of the electrons were opposite to the magnetic field. But since this
is a higher energy state that is less probable. To enhance this state,
two solutions exist. The first is to set the temperature to an energy
where the upper energy state would be more probable. The second solution
is to set the the atom in an external magnetic quadrupole. By placing
the nuclide just outside the center, the electron would set the lower
state in the opposite magnetic field in the far point, and be in the
higher energy state when close to the nuclide. Another condition for
energy extraction by magnetic quadrupole polarizability is that the
electron must be in a single state to be able to spin flip. Suitable
elements that are naturally in 1S states are the alkali and coin metals.
In addition, other metals could be in 1S states, including nickel,
platinum, niobium, molybdenum, ruthenium, rhodium, and chromium.

\subsection*{Comparision between theory and experiment}

The experiment described in appendix has both weak and strong links
to the theory. The weaker links are basically the atomic source theory.
Since no detailed study in varying the atomic structure properties
has been carried out, only arguments drawn by the electrical current, and chemical
composition may be drawn. 

The single S-state condition motivates the presence of the elements
nickel, lithium and hydrogen. The current solution of$\left(\bar{E}\times\dot{\bar{E}}\right)$
matches the required presence of current in the experiment. It also
matches the charge-neutral plasma observation. This is because the
$\sigma\cdot\left(\bar{E}\times\dot{\bar{E}}\right)$ condition would
be positive at half the time if $\sigma$ alignment was random thus
enhancing 50\% of the current to induce kinetic energy for the heavy
ions. The stronger links are found for the isotopic shifts. The proposed
mechanism has already been presented in the long-range strong force
part. 

Also the existance of a plasma indicates that the positive ions should
have a sizeable kinetic energy to overcome the air gap barrier in
the starting phase. This requires a non zero part of the nuclide kinetic
energy in \ref{eq:erel} and motivates the existance of the $\sigma_{I=2}$
potentiall.

\section*{Summary and discussion}

The aim of the theoretical parts of this paper was to answer two questions: 
\begin{itemize}
\item How to extract energy without strong radiation.
\item How to effect this over a long range. 
\end{itemize}
The answer was done in brief, first comparing photon-meson interaction
in polarizability with nucleon-nucleon interaction; secondly to use
a process similar to the Primakoff effect with a $\sigma$ meson replacing
the pion. The advantage with the $\sigma$ meson over the pion is
that the scalar meson is included in the strongest part of the strong
force. The idea is no stranger than that of Cooper pairs in superconductivity,
where a pair of electrons, which initially would repel each other,
display an attractive interaction due to the special EM-field conditions
that are created by the lattice between them. 

For the polarizability part, ranges where the free nucleon is transformed
into a special state with the energy of a bound one were found by searching
for the condition $H_{eff}>0$ for theoretical and experimental polarizability
constants. The polarizability interaction only answers the first question
and for the second question two special electron nucleon interaction
where introduced. Either the special EM fields could create an enviroment
where the electron could carry a nucleon or one where the electron
triggered a special long range potential of the strong force from
the nucleon. The second solution is favourable due to the fact that
the first would still create some radiation and hence not be able
to answer the first question.

\paragraph*{Discussion}

The polarizabilty parameters needed are not known from experiment
except for the spin polarizability constants. The long-range strong
force that arises from the $\sigma_{I=2}$ meson is an interesting
new part of the strong force that seems to be necessary to explain
the experimental data but is not completely needed for the polarizability
part. This long range potential is also unknown, both in detailed
theory and experiment. To extract those constants experimentally,
a theoretical means would be to use $\pi\pi$-lepton scattering with
a measurement of nucleon properties in a nearby region. Practically,
whether a pion beam with high enough luminosity as well as the means
to construct it are possible is questionable. It is also plausible
that the isospin mixing of the $\sigma$ is a property only found
in baryons, as a cause or a consequence of baryon number conservation.

Open questions that calls for future experimentation are:
\begin{itemize}
\item Is the main source spin polarizability, magnetic quadrupole polarizability
or both?
\item Is the long-range problem solved by e-N meson interaction or $\sigma_{I=2}$
enhanced exchange potential?
\item What is the neccesary atomic states for creating the special EM fields?
\item What is the relation $E_{k}\left(e^{-}\right)$ vs. $\sum E_{k}\left(N\right)$?
\end{itemize}

Experimental methods that could be used to solve those questions are:
\begin{itemize}
\item An isotopic analysis of silicon and chromium isotopes in the ash,
in order to obtain a solid confirmation of proton transfer.
\item Measuring helium output.
\item Energy output variation by changing current and pressure. 
\item A beta spectrum probed inside the nickel. 
\item More probable chemical elements in the fuel for example all alkli
metals.
\end{itemize}

\section*{Appendix}

\subsection*{Experiment }

The experimental inputs have two sources: one is isotopic shifts observed
in the experiments performed in Lugano\cite{Lugano} and by Parkhomov\cite{Parkhomov};
the other source is the observation of an energy emitting charge-neutral
plasma with current running through it observed in two different experimental
setups, one with energy measured from the spectral maximum of the
plasma and the other from a fluid heat exchange measuring system.

\subsubsection*{Isotopic shifts}

Starting with the isotopic shifts, what is noteworthy is that most
stable nucleons have a ground state of $0^{+}$, and that all seen
final state nucleons are of this state. The main reaction observed
are: 

$^{62-x}Ni+xn^{*}\to^{62}Ni$

$^{27}Al+p^{*}\to^{28}Si$

where $p^{*}$ and $n^{*}$ mean a bound nucleon, the source of which
is in another nuclide. Noteable are that the $^{27}Al+p^{*}$ reaction
only is validated by chemical observation. Other possible reactions
that would be feasable within the same theory are: 

$^{58}Ni+2n^{*}\to^{60}Ni$

$^{7}Li+p^{*}\to^{8}Be\to2\alpha$

The main sources of the bound nucleons are:

$^{64}Ni\to^{62}Ni+2n^{*}$

$^{7}Li\to^{6}Li+n^{*}$

$^{7}Li\to\,^{6}He+p^{*}$

$^{55}Mn\to^{54}Cr+p^{*}$

The last reaction has only been observed by chemical identification.

\subsubsection*{Experiment with energy measurement done with a spectrometer}

Description of the apparatus: 

The circuit of the apparatus consists of a power source supplying
direct current, a 1-Ohm resistor load, and a reactor containing two
nickel rods with LiAlH4 separated by 1.5 cm of space.

Measurements: 

During the test, a direct current was switched on and off. When the
current was switched on, a plasma was seen flowing between the two
nickel rods. The current was running through the plasma but the plasma
was found to be charge-neutral from a Van de Graaff test. This implies
that the plasma has an equal amount of positive ions flying in the
direction of the current and negative ions(electrons) in the opposite
direction.

Input: 0.105 V of direct current over a 1 Ohm resistance.

Energy output: The wavelength of the radiations from the reactor was
measured with a spectrometer ( Stellar Net spectrometer 350-1150 nm
) and was integrated with the value of 1100 nm ( 1.1 microns ). 

The temperature of the surface of the reactor ( a perfect black body
) was calculated with Wien's equation: 2900/$\lambda$ (micron) =
2900/1.1 = 2636 K 

As per Boltzmann\textquoteright{}s equation, the effect is: $W=\sigma\times\epsilon\times T^{4}\times A$

A = 1.0 $cm^{2}$ 

$\epsilon$ = 0.9 

By substitution: $W=5,67\times10^{12}\times0.9\times4.8\times10^{13}=244.9$

\subsubsection*{Experiment with energy measurement done with a heat exchanger}

\begin{figure}
\includegraphics[angle=270,scale=0.15]{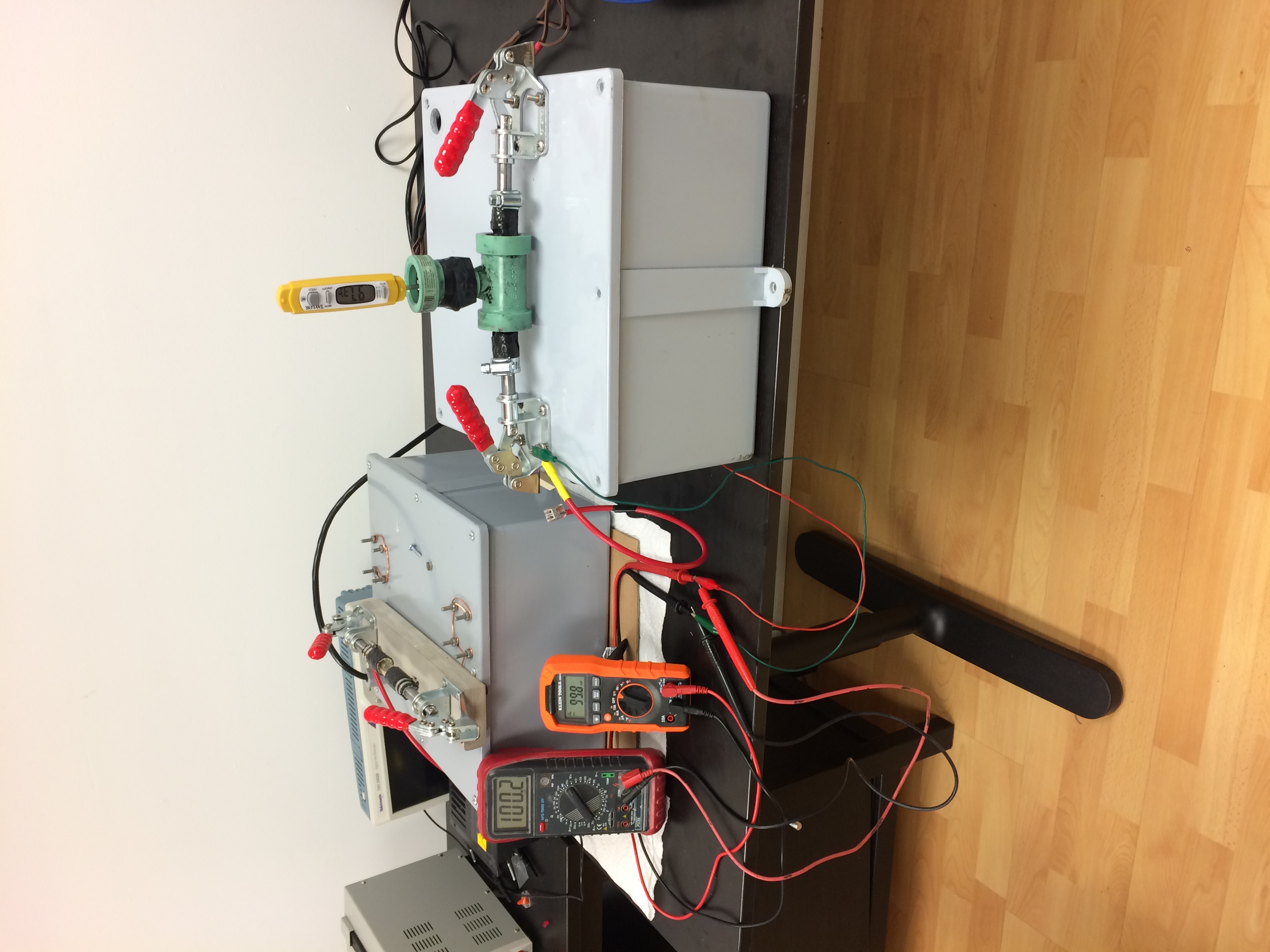}

\caption{\label{fig:exp-2}Experimental setup with energy measured by a heat
exchanger.}
\end{figure}

 The system is displayed in figure \ref{fig:exp-2}. In the figure,
the yellow thermometer measures the temperature of the oil inside
the heat exchanger. In the left in the figure there are two voltmeters
that measure the mV of the current passing through the 1 Ohm brown
resistance. In order to not burn the equipmet, the experiment was set in a state with lower output power. 

Calculations of the calorimetry made by the heat exchanger: 

efficiency of the heat exchanger:90\%

Primary heat exchange fluid: lubricant oil ( Shell mineral oil ) 

Characteristics of the lubricant oil: D = 0.9 Specific Heat: 0.5 

Calorimetric data of the fluid: 0,5 Kcal/h = 0.57 Wh/h 

Flow heating: 1.58 C / 1.8\textquotedbl{} x 11 g 

Resulting rating: 20 Wh/h 

Energy input: V=0.1 R=1 Ohm $\to$ W=0.01

\newpage{}

\end{document}